\documentclass[aps,prd,tightenlines,preprint,nofootinbib,a4paper]{revtex4}
\usepackage{epsfig}
\usepackage{makecell}

\begin{document}

\title{Rephasing invariance and permutation symmetry in flavor physics}

\author{T. K. Kuo\footnote{tkkuo@purdue.edu}}
\affiliation{Department of Physics, Purdue University, West Lafayette, IN 47907, USA}

\author{S. H. Chiu\footnote{schiu@mail.cgu.edu.tw}}
\affiliation{Physics Group, CGE, Chang Gung University, 
Taoyuan 33302, Taiwan}

\begin{abstract}

With some modifications, the arguments for rephasing invariance can be used to establish
permutation symmetry for the standard model. The laws of evolution of physical variables,
which transform as tensors under permutation, are found to obey the symmetry, explicitly.
We also propose to use a set of four mixing parameters, with unique properties,
which may serve to characterize flavor mixing. 

\end{abstract}


\maketitle

\pagenumbering{arabic}



\section{Introduction}

One of the long-standing puzzles in the Standard Model (SM) is the existence of three 
families of fundamental fermions. (Throughout this paper, SM refers to a modified version
with the addition of three massive Dirac neutrinos, so that one may treat the lepton
sector on a par with the quark sector.) These fermions are endowed with properties
(masses and mixing parameters) which seem to be arbitrary.  At the same time,
it is also well-established that these properties are not static, but that they do
evolve as the physical environment changes. They may thus be regarded as dynamical variables in the system. E.g., the masses and mixing parameters are dependent on the energy scale, and evolve according to the renormalization group equations (RGE).
Similarly, as a coherent neutrino beam propagates, the neutrino mixing parameters
change along its path. A third physical example concerns neutrino propagation in matter,
where one finds that their masses and mixing become functions of the medium density.

While it may be difficult to find order in the observed fermion properties,
one can look for regularity in the laws of their evolution. In a previous 
paper \cite{Kuo:2018mnm}, it was pointed out that these equations obey the  
symmetry $[S_{3}]^{4}=(S_{3}(u),S_{3}(d),S_{3}(\ell),S_{3}(\nu))$,
where $S_{3}$ is the permutation group which operates on the three members in each
of the four fermion sectors.  Examples of calculated formulas of transition
(neutrino oscillation in vacuum), as well as evolution equations (RGE and
neutrino oscillation in matter) were examined. It was found that they obey
the permutation symmetry, explicitly.

In this paper, we will incorporate, systematically, rephasing into our study of permutation.
It turns out that the two operations are closely related.  Let $V_{\alpha i}$ denote
the elements of either the CKM or the PMNS matrix. While rephasing attaches phases to
$V_{\alpha i}$ according to their indices, permutation exchanges these indices.
We will demonstrate that the SM Lagrangian, $\mathcal{L}_{SM}$, is invariant in
form under either operation.  Furthermore, rephasing invariance implies that $V_{\alpha i}$
must be grouped in specific combinations according to their indices.
Applying a permutation to any such product is seen to yield another rephasing invariant
combination. This means that any physical variable constructed out of $V_{\alpha i}$
must belong to a tensor under $S_{3}$.  Also, its evolution equation obeys the
permutation symmetry.

The use of permutation tensors for physical variables has another property, 
owing to the few available representations of $S_{3}$.
This prevents the proliferation of physical variables.  Indeed, we find that there are many
relations between possible construction of rephasing invariant combinations (RIC),
so that only a few are independent.

Permutation considerations can also shed light on another problem in flavor physics.
To parametrize the mixing of flavors, it seems natural to use $V_{\alpha i}$,
or $W_{\alpha i}=|V_{\alpha i}|^{2}$, as variables. However, there is no good
criterion to pare down the set to four physical ones. We will identify a set of
four variables, all of which transform as singlets under permutation, and can be
used as physical parameters.  Some of their properties, including a set of
RGE, are presented.

This paper is organized as follows. 
In Sec. II we briefly introduce the notations and the physical variables
which have simple properties under permutation. A systematic analysis of rephasing
and permutation is given in Sec. III. Sec. IV provides examples that reveal
permutation symmetry in some known results, such as neutrino oscillation
in vacuum and in matter, and the RGE for quarks. In Sec. V, a set of 
four variables, $\{D,Q^{2},J^{2},K\}$, are proposed as alternatives that
facilitate the study of flavor physics. We then summarize this work in Sec. VI.

\section{Notation and Mathematical Preliminaries}

In order to facilitate the study of rephasing invariance and permutation,
it is important to choose variables which have simple properties under their operations.
Let us start with $V_{\alpha i}$ (To avoid repetition, we will use $V_{\alpha i}$ to denote
elements of the PMNS matrix.  The case for CKM is the same.) We will impose,
without loss of generality, the condition \cite{Kuo:2005pf}    
\begin{equation}
det V=+1,
\end{equation}
so that
\begin{equation}\label{II2}
V^{*}_{\alpha i}=\frac{1}{2!}e_{\alpha \beta \gamma}e_{ijk}V_{\beta j}V_{\gamma k},
\end{equation}
and rephasing is given by
\begin{equation}
V_{\alpha i} \rightarrow \mbox{e}^{i\phi_{\alpha}}V_{\alpha i}\mbox{e}^{-i\phi_{i}},
\end{equation}
with $\sum \phi_{\alpha}=\sum \phi_{i}=0$.
From Eq. (\ref{II2}), we may eliminate $V_{\beta j}^{*}$ from any expression
in favor of only products of the $V_{\alpha i}$'s.  
One can then construct basic rephasing invariant combinations (RIC) in the form
(no summation over capital indices)
\begin{equation}\label{II4}
\Gamma^{AB\Gamma}_{IJK}=E_{AB\Gamma}E_{IJK}V_{AI}V_{BJ}V_{\Gamma K},
\end{equation}
where $E_{\alpha \beta \gamma}$ 
is the symmetric Levi-Civita symbol \cite{Kuo:2018mnm},  
which is symmetric under exchange, and
\begin{eqnarray}
E_{\alpha \beta \gamma}&=&1, \alpha \neq \beta \neq \gamma ;  \nonumber \\ 
                                      &=& 0, \mbox{otherwise} \nonumber
 \end{eqnarray}
Thus, $\Gamma$ contains the index of each row (and each column) once, 
and only once.

It is also seen that the difference of two $\Gamma$'s is equal to some 
$|V_{\alpha i}|^{2}=W_{\alpha i}$, so that all the $\Gamma$'s must have the same
imaginary part, which can be identified with $J$, the Jarlskog invariant \cite{Jarlskog:1985ht}.  
We write
\begin{equation}
\Gamma^{e\mu \tau}_{IJK}=R_{IJK}-iJ,
\end{equation}
with the definition 
$(R_{123},R_{231},R_{312};R_{132},R_{213},R_{321})=(x_{1},x_{2},x_{3};y_{1},y_{2},y_{3})$,
which satisfy the consistency conditions
\begin{equation}\label{con6}
\sum x_{i}-\sum y_{i}=1,
\end{equation}
\begin{equation}\label{con7}
\sum_{i>j}x_{i}x_{j}=\sum_{i>j}y_{i}y_{j}.
\end{equation}

We now turn to the properties of $V_{\alpha i}$ under permutation.
Under an exchange, $(ij)$, we have
$V_{\alpha i} \leftrightarrow (\mbox{phase})V_{\alpha j}$, where an arbitrary phase
is associated with the exchange operator.  In order to maintain $det V=+1$, we choose
\begin{equation}
(V_{\alpha i},V_{\alpha k}) \leftrightarrow -(\mbox{phase})' (V_{\alpha j},V_{\alpha k}),
\end{equation}
where further possible rephasing is contained in (phase)$'$.
However, if we include $V_{\alpha i}$ in an RIC, these phases cancel out and we have
\begin{equation}\label{II9}
(V_{\alpha i},V_{\alpha k}) \longleftrightarrow -(V_{\alpha j},V_{\alpha k}) (\mbox{in rephasing invariant combinations}).
\end{equation}
This is the behavior of $V_{\alpha j}$ under the exchange $(ij)$.
Similarly, under 
$(\alpha \beta)$, $(V_{\alpha i},V_{\gamma i}) \leftrightarrow -(V_{\beta i},V_{\gamma i})$, etc. 
Thus, in the notation of Ref.\cite{Kuo:2018mnm},  
$V_{\alpha i} \sim \widetilde{\bf{3}} \times \widetilde{\bf{3}}$ (also, same for $V^{*}_{\alpha i}$),
provided that $V_{\alpha i}$ is included in an RIC.

In analogy to $V^{*}_{\alpha i}$ (Eq. (\ref{II2})), we can define
\begin{equation}
V'_{\alpha i}=\frac{1}{2!}E_{\alpha \beta \gamma}E_{ijk}V_{\beta j}V_{\gamma k},
\end{equation}
which transforms as $\bf{3} \times \bf{3}$. Now we have
\begin{equation}\label{eq11}
\frac{1}{2!}e_{A\beta\gamma}e_{Ijk}V_{AI}V_{\beta j}V_{\gamma k}=|V_{AI}|^{2}=W_{AI},
\end{equation}
\begin{equation}
\frac{1}{2!}E_{A\beta \gamma}E_{Ijk}V_{AI}V_{\beta j}V_{\gamma k}=V'_{AI}V_{AI}
=w_{AI}-2iJ.
\end{equation}
These are identified with the variables ($W_{AI},w_{AI}$) 
introduced earlier \cite{Kuo:2005pf}, 
and we can verify
\begin{equation}
w_{AI}=\frac{1}{2!}e_{A\beta \gamma}e_{Ijk}W_{\beta j}W_{\gamma k}
\end{equation}
by using properties of the $(x_{i},y_{j})$ variables. Also,
\begin{equation}
\frac{1}{2!}e_{A\beta \gamma}e_{Ijk}w_{\beta j}w_{\gamma k}= D W_{AI},
\end{equation}
\begin{equation}
D=\sum_{i}w_{\alpha i}=\sum_{\alpha}w_{\alpha i}=det W.
\end{equation}
The relation,
\begin{equation}\label{eq16}
\frac{1}{3!}E_{\alpha \beta \gamma}E_{ijk}V_{\alpha i}V_{\beta j}V_{\gamma k}=D-i(3!)J,
\end{equation}
shows that $D \sim \widetilde{\bf{1}} \times \widetilde{\bf{1}}$ and 
$J \sim \widetilde{\bf{1}} \times \widetilde{\bf{1}}$. While the transformation
property of $D$ is expected, that $J$ behaves like a pseudo-P-scalar is very interesting.
We will return to this point in our analysis later.

It is useful to introduce, explicitly, a set of $3 \times 3$ matrices which are representations of
$S_{3}$ (with elements ($e, (1,2,3),(1,3,2);(3,2),(2,1),(1,3)$):
\begin{eqnarray}\label{eq17}
&X^{0}_{1}&=
\left(\begin{array}{ccc}
   1& 0 & 0 \\
   0 & 1 & 0 \\
   0 & 0 & 1 \\
    \end{array}
    \right), 
  X^{0}_{2}=
\left(\begin{array}{ccc}
   0& 1 & 0 \\
   0 & 0 & 1 \\
   1 & 0 & 0 \\
    \end{array}
    \right),  
    X^{0}_{3}=
\left(\begin{array}{ccc}
   0& 0 & 1 \\
   1 & 0 & 0 \\
   0 & 1 & 0 \\
    \end{array}
    \right);  \nonumber \\
 &X'_{1}&=
\left(\begin{array}{ccc}
  1& 0 & 0 \\
   0 & 0 & 1 \\
   0 & 1 & 0 \\  
    \end{array}
    \right),
  X'_{2}=
\left(\begin{array}{ccc}
   0& 1 & 0 \\
   1 & 0 & 0 \\
   0 & 0 & 1 \\
    \end{array}
    \right),  
    X'_{3}=
\left(\begin{array}{ccc}
   0& 0 & 1 \\
   0 & 1 & 0 \\
   1 & 0 & 0 \\
    \end{array}
    \right).    
\end{eqnarray}

While the set $\{[X]\}=\{[X^{0}_{i}],[X'_{i}]\}$ represents $S_{3}$ on ${\bf{3}}$, another set
\begin{equation}
\{[\widetilde{X}]\}=\{[X^{0}_{i}],-[X'_{i}]\}
\end{equation}
is for $\widetilde{\bf{3}}$.  Thus, if we write $V_{\alpha i}$ as a matrix, $[V]$,
a permutation of the index $i$ is given by 
\begin{equation}
[V] \rightarrow [V][\widetilde{X}]^{\dag},
\end{equation}
while that on the $\alpha$ index is given by left multiplication
\begin{equation}
[V] \rightarrow [\widetilde{X}][V].
\end{equation}

The matrices for $W_{\alpha i}$ and $w_{\alpha i}$ can be written as
\begin{equation}
[W]=\sum x_{i}[X^{0}_{i}]-\sum y_{i}[X'_{i}],
\end{equation}
\begin{equation}
[w]=\sum x_{i}[X^{0}_{i}]+\sum y_{i}[X'_{i}].
\end{equation}
Permutations on the index $i$ are given by
\begin{equation}
[W] \rightarrow [W] [X]^{\dag},
\end{equation}
\begin{equation}
[w] \rightarrow [w][\widetilde{X}]^{\dag},
\end{equation}
while for index $\alpha$ one would have $[X]$ and $[\widetilde{X}]$   
multiplying from the left.

It follows that under either $(i j)$ or $(\alpha \beta)$,
$x_{l} \leftrightarrow -y_{m}$ for appropriate $l$ and $m$. Also,
$\sum x_{i} \leftrightarrow -\sum y_{i}$, from $(ij)$ or $(\alpha \beta)$.
Thus, $D=\sum(x_{i}+y_{i})$ is a pseudo-P-scalar, as before.
We also find $Q^{2}=\sum x_{i}x_{j}+\sum y_{i}y_{j} \rightarrow Q^{2}$, and
$J^{2}=\Pi x_{i}-\Pi y_{i} \rightarrow J^{2}$, so that both are P-scalars.

Another familiar 
RIC \cite{Jarlskog:1985ht,Chiu:2015ega} is, for $(A \neq B \neq \Gamma,I \neq J \neq K)$, 
\begin{equation}\label{II25}
\Lambda_{AI}=Re(\Pi^{B\Gamma}_{JK})=
Re(V_{BJ}V_{\Gamma K}V^{*}_{BK}V^{*}_{\Gamma J}).
\end{equation}
Under permutation, it transforms $\sim \bf{3} \times \bf{3}$.
It turns out that we can express $\Lambda_{AI}$ in terms of $W$,
\begin{equation}\label{eq26}
2\Lambda_{AI}=\frac{1}{2}E_{A\beta\gamma}E_{Ijk}W_{\beta j}W_{\gamma k}-W_{AI}.
\end{equation}
A similar construction using $w$ is
\begin{equation}\label{eq27}
2\Lambda'_{AI}=\frac{1}{2}E_{A\beta \gamma}E_{Ijk}w_{\beta j}w_{\gamma k}-D w_{AI}.
\end{equation}

Some other tensors will also appear in RGE calculations.  There is a tensor
$(\sim \widetilde{{\bf{3}}}\times \widetilde{{\bf{3}}} + 
\widetilde{{\bf{1}}}\times \widetilde{{\bf{1}}})$
which was discussed earlier \cite{Chiu:2015ega}: 
\begin{eqnarray}\label{eq28}
\Xi^{0}_{AI}&=& E_{AB\Gamma}E_{IJK}V_{AJ}V_{AK}V^{*}_{AI}V_{BI}V_{\Gamma I} \nonumber \\
&=&(Re \Xi^{0})_{AI}+iJ(1-W_{AI}),
\end{eqnarray}
\begin{equation}
-[Re \Xi^{0}]=\left(\begin{array}{ccc}
   x_{2}x_{3}-y_{2}y_{3}& x_{3}x_{1}-y_{3}y_{1} & x_{1}x_{2}-y_{1}y_{2} \\
   x_{1}x_{2}-y_{3}y_{1}& x_{2}x_{3}-y_{1}y_{2} & x_{3}x_{1}-y_{2}y_{3} \\
   x_{3}x_{1}-y_{1}y_{2} & x_{1}x_{2}-y_{2}y_{3} & x_{2}x_{3}-y_{3}y_{1} \\
    \end{array}
    \right)
\end{equation}
Another useful tensor 
($\sim \bf{3} \times \bf{3}+\bf{1} \times \bf{1}$) 
was also used \cite{Chiu:2015ega}, 
\begin{eqnarray}
Z^{0}_{AI}&=& E_{AB\Gamma}E_{IJK}V_{AJ}V_{AK}V'_{AI}V_{BI}V_{\Gamma I}+
2J^{2} \nonumber \\
&=&[Re Z^{0}]_{AI}-iJ(D-w_{AI}),
\end{eqnarray}
\begin{equation}
[Re Z^{0}]=\frac{1}{2!}\sum_{i\neq j \neq k}(x_{j}x_{k}[X^{0}_{i}]+y_{j}y_{k}[X'_{i}]).
\end{equation}
Finally, one also needs the tensor $(\bf{3} \times \bf{3})$,
\begin{equation}\label{eq32} 
w^{2}_{AI}-W^{2}_{AI}=2(\Lambda_{AI}-\Lambda'_{AI}),
\end{equation}
in addition to the identity
\begin{equation}\label{eq33} 
2[Re Z^{0}]_{AI}-Q^{2}=\Lambda_{AI}+\Lambda'_{AI}.
\end{equation}
These relations and other identities will be further discussed in Appendix A.

\section{Rephasing Invariance and Permutation Symmetry}

We now turn to a systematic study of rephasing and permutation.
We concentrate on the SM in a minimally extended version, with massive Dirac neutrinos.
For our purposes only the EW interactions need to be considered, so that
we will only study the lepton sector explicitly, bringing in the parallel quark sector
when appropriate. In this case, of the many parts of the SM Lagrangian, $\mathcal{L}_{SM}$,
we can focus on the leptonic EW interactions in the mass eigenstate basis.
Schematically, we write
\begin{eqnarray}\label{3.1}
\mathcal{L}_{SM}(\ell,H) &\sim& \sum_{\alpha,i}[\bar{\Psi}_{\alpha}V_{\alpha i}\psi_{i}-(1+\frac{h}{v})(m_{\alpha}\bar{\Psi}_{\alpha}\Psi_{\alpha}+m_{i}\bar{\psi}_{i}\psi_{i})]+ \cdot \cdot \cdot \\ \nonumber
&=&[\bar{\Psi}]^{T}[V][\psi]-(1+\frac{h}{v})([\bar{\Psi}]^{T}[M_{\ell}][\Psi]+
[\bar{\psi}]^{T}[M_{\nu}][\psi])+ \cdot \cdot \cdot
\end{eqnarray}
Here, to simplify the notation, we omit the gauge fields ($W_{\mu}$, in 
$J_{\mu}W_{\mu}^{\dag}+h.c.$) and proper Dirac matrices. Also,
$\alpha=(e,\mu,\tau)$, $\psi_{i}$ refers to $\nu_{i}$, and $m_{\alpha} (m_{i})$
are their masses. $h$ denotes the Higgs field in the physical gauge, $v$ is the VEV,
and $V_{\alpha i}$ is an element of the PMNS matrix.  In an obvious matrix notation,
$[\bar{\Psi}]^{T}=(\bar{\Psi}_{e},\bar{\Psi}_{\mu},\bar{\Psi}_{\tau})$, etc., and
the diagonal mass matrices are $[M_{\ell}]$ and $[M_{\nu}]$.

Eq.~(\ref{3.1}) is the result of diagonalizing the Higgs-Fermion coupling by U(3)
transformations on $\Psi_{\alpha}$ and $\psi_{i}$. However, this procedure is not unique.
A familiar example is the rephasing transformation:
\begin{equation}
[\Psi] \longrightarrow P[\Psi],
\end{equation}
\begin{equation}
[\psi] \longrightarrow P'[\psi],
\end{equation}
where $(P,P')$ are diagonal phase matrices which, to maintain $detV=+1$,
satisfy $detP=detP'=+1$.  The Lagrangian is invariant in form provided that $V$
also changes according to
\begin{equation}
[V] \longrightarrow P[V]P'^{\dag}.
\end{equation}

Similarly, there is another transformation on $[\Psi]$ and $[\psi]$ which leaves
$\mathcal{L}_{SM}$ invariant in form:
\begin{equation}
[\Psi]\longrightarrow [\widetilde{X}][\Psi].
\end{equation}
\begin{equation}
[\psi] \longrightarrow [\widetilde{X}'][\psi],
\end{equation}
\begin{equation}\label{III7}
[V] \longrightarrow [\widetilde{X}][V][\widetilde{X}']^{\dag},
\end{equation}
\begin{equation}\label{III8}
[M_{\ell}] \longrightarrow [\widetilde{X}][M_{\ell}][\widetilde{X}]^{\dag},
\end{equation}
\begin{equation}\label{eq42}
[M_{\nu}] \longrightarrow [\widetilde{X}'][M_{\nu}][\widetilde{X}']^{\dag}.
\end{equation}
Note that $[\widetilde{X}]$ and $[\widetilde{X}']$ are used here in accordance with 
Eq.~(\ref{II9}).  Also, the effects of Eqs.(\ref{III8}-\ref{eq42}) are to keep $[M_{\ell}]$ and 
$[M_{\nu}]$ diagonal, but to reshuffle their matrix elements. I.e., both $m_{\alpha}$
and $m_{i}$ transform as $\bf{3}$. 
The upshot of our analysis is that, with the assignments
$(\Psi,\psi) \sim \widetilde{\bf{3}}$, $V \sim \widetilde{\bf{3}} \times \widetilde{\bf{3}}$,
and $(m_{\alpha},m_{i}) \sim (\bf{3}_{\ell},\bf{3}_{\nu})$, $\mathcal{L}_{SM}(\ell)$ has an exact symmetry,
$S_{3}(\ell) \times S_{3}(\nu)$. Including the quark sector, the full $\mathcal{L}_{SM}$
is symmetric under $[S_{3}]^{4}=S_{3}(\ell)\times S_{3}(\nu) \times S_{3}(u) \times S_{3}(d)$.

It is noteworthy that rephasing invariance and permutation symmetry are so closely
related. While rephasing can be balanced out by a corresponding operation on $V$, 
a permutation of the wave functions can be countered with similar actions on $V$
and on the masses.  Among the possible U(3) operation on the wave functions, rephasing
and permutation are unique since any unitary transformation on a matrix
cannot change its eigenvalues, except possibly their ordering.

As noted before, the basic RIC, given in Eq.~(\ref{II4}), contains each family
index once, and only once. A permutation of indices on any RIC yields another RIC.
Since physical variables are composed of products of RIC's, starting from any variable,
repeated permutations generates a tensor.  That is, physical variables are naturally
grouped into tensors under $S_{3} \times S_{3}$.
Coupled with the invariance of
$\mathcal{L}_{SM}$, this means that the evolution of these tensors is regulated by the permutation symmetry.  As we will demonstrate explicitly in the next section, there are many
known examples which exhibit the symmetry.  Considerations on permutation
offers interesting insights into these equations, they can also serve as useful tools
to check the validity of future calculations in flavor physics.


\section{Examples}

In this section, we investigate the implications of the permutation symmetry as 
applied to some known results in flavor physics. At first sight, a permutation,
such as $(\psi_{i},V_{\alpha i},m_{i}) \rightarrow (\psi_{j},V_{\alpha j},m_{j})$,
seems rather innocuous and inconsequential.  Indeed, if one considers, e.g., 
the decays $b \rightarrow u + \ell \nu$ and $b \rightarrow c+ \ell \nu$, at the tree level,
it is obvious that the two procedures are related by the permutation
$(u,V_{ub}) \rightarrow (c,V_{cb})$. However, a more interesting situation arises when a calculation involves internal fermions.  In this case, the vestige of their participation
is contained in a function $f(V_{\alpha i},m_{i})$, corresponding to using $\psi_{i}$,
in a certain order, as the basis of calculation.  Had one chosen the basis
$[\widetilde{X}]\psi_{i}$, one would have obtained the function $f$ with permuted indices.
With permutation symmetry, this implies that $f$ must be a function of
permutation invariants, such as $f(\sum m_{i}W_{\alpha i})$. In the following
we will revisit some examples that were studied 
before \cite{Kuo:2018mnm}, 
with further comments. Additional examples will also be given.

\subsection{Neutrino oscillation in vacuum}

The probability function for neutrino oscillation is well-known. In tensor notation, 
it can be written \cite{Kuo:2018mnm} as (for $\alpha \neq \beta$)  
\begin{equation}
P(\nu_{\alpha} \rightarrow \nu_{\beta})=-4E_{\alpha \beta \gamma}\delta_{ij}
\Lambda_{\gamma i}\sin^{2}\widetilde{\Phi}_{j}+2J(\sum \sin 2\widetilde{\Phi}_{i}),
\end{equation}
Here, $\Lambda_{\gamma i}$ is defined in Eq.(\ref{II25}).   
$\widetilde{\Phi}_{i}(\sim \widetilde{\bf{3}})$ is a phase factor given by
$\widetilde{\Phi}_{i}=\frac{1}{2}e_{ijk}(m^{2}_{j}-m^{2}_{k})(L/4E)$.
This result is obtained by considering the propagation (internally)
of the mass eigenstate, $|\nu_{i}\rangle$, from $t=0$ to $t=L$.
Had we used an equivalent but permuted basis, $\widetilde{X}|\nu_{i}\rangle$,
we would have obtained the same probability function, only with the indices permuted.
The $S_{3}(\nu)$ symmetry then dictates that the function must be a P-scalar,
which is indeed the case.  Also as we have emphasized before,
the permutation property of $J$, that $J \rightarrow -J$ under any exchange 
$\alpha \leftrightarrow \beta$ (and $i \leftrightarrow j$), implies that the second term in
$P(\nu_{\alpha}\rightarrow \nu_{\beta})$ is T (and CP) violating, from symmetry arguments without the need
to do any calculation.

\subsection{Neutrino oscillation in matter}

When a neutrino passes through a medium, an effective mass for $\nu_{e}$
is generated.  This produces changes in the neutrino parameters. 
For an infinitesimal changes, $dA$, $A=2\sqrt{2} G_{F}N_{e}E$, with 
$(\delta H^{D})_{\xi}=(dA,0,0)=(\delta H^{D}_{ee},0,0)$, the induced changes in neutrino parameters are given by (\cite{Kuo:2018mnm,Chiu:2017ckv}, 
see also \cite{Xing:2018lob}, for an analysis in the PDG variables)  
\begin{equation}
\delta m_{i}^{2}=(\delta H^{D})_{\xi}W_{\xi i},
\end{equation}
\begin{equation}
\delta W_{\alpha i}=2E_{\alpha \beta \gamma}e_{ijk}(\delta H^{D})_{\beta}
\Lambda_{\gamma j}/\Delta \widetilde{D}_{k},
\end{equation}
\begin{equation}
\delta(\ln J)=-(\delta H^{D})_{\xi}\Delta \widetilde{W}_{\xi k}/\Delta \widetilde{D}_{k},
\end{equation}
where $\Delta \widetilde{D}_{k}=(\frac{1}{2})e_{klm}(D_{l}-D_{m})$, 
$\Delta \widetilde{W}_{\xi k}=(\frac{1}{2})e_{klm}(W_{\xi l}-W_{\xi m})$.
These equations are invariant in form under $S_{3}(\ell) \times S_{3}(\nu)$.
To arrive at these results, one chooses a certain basis $|\nu_{i} \rangle$.
However, one could have chosen to use $\widetilde{X}|\nu_{i}\rangle$ as basis, leading to the same physics. This freedom of choice is reflected in the symmetric tensor forms of these equations,

It is also interesting to note that there are two ``matter invariants" obtained a 
long time ago  \cite{Naumov:1993vz,Harrison:1999df,Kimura:2002hb,Toshev:1991ku,Krastev:1988yu,Chiu:2010da}. 
They can be written in the tensor notation 
(see Eqs. (32-33) in Ref. \cite{Chiu:2017ckv})  
\begin{equation}
\frac{d}{dA} \ln[J(D_{1}-D_{2})(D_{2}-D_{3})(D_{3}-D_{1})]=0,
\end{equation}
\begin{equation}
\frac{d}{dA}[J^{2}/W_{e1}W_{e2}W_{e3}]=0.
\end{equation}
It is seen that both expressions are invariant under $S_{3}(\nu)$.
Moreover, note the pairing of $J (\sim \widetilde{\bf{1}})$ with
$E_{ijk}\widetilde{D}_{i}\widetilde{D}_{j}\widetilde{D}_{k}$, which is another
pseudo-P-scalar, and that of $J^{2}(\sim \bf{1})$ with $E_{ijk}W_{ei}W_{ej}W_{ek}$.

\subsection{One-loop RGE for quarks}

The evolution of the physical parameters of fermions has been studied for a long time.
We will discuss here only the RGE for quarks, since with Dirac neutrinos,
the leptons behave just like the quarks.  Traditionally, these equations were given using
the PDG parameters.  The results are rather complicated 
(e.g., Ref.\cite{Balzereit:1998id}). 
In terms of the tensor notation, the equations for the mixing parameters can be
written in the form \cite{Kuo:2018mnm}  

\begin{equation}\label{IV7}
\mathcal{D}W_{\alpha i}=-2c' \cdot (\Delta \widetilde{m}^{2}_{\beta}[S^{\alpha i}]_{\beta j}\widetilde{G}_{j}+
\Delta \widetilde{m}^{2}_{j}[S^{\alpha i}]^{T}_{j\beta}\widetilde{H}_{\beta})
\end{equation}

\begin{equation}\label{IV8}
\mathcal{D}\ln J=-c'(\Delta \widetilde{m}^{2}_{\alpha}
w_{\alpha i}\widetilde{G}_{i}+\Delta \widetilde{m}^{2}_{i}
w_{i \alpha}^{T}\widetilde{H}_{\alpha}).
\end{equation}
Here, $\Delta \widetilde{m}^{2}_{\alpha}=(\frac{1}{2})e_{\alpha \beta \gamma}(m^{2}_{\beta}-m^{2}_{\gamma})$,
$\widetilde{H}_{\alpha}=(\frac{1}{2})e_{\alpha \beta \gamma}(m^{2}_{\beta}+m^{2}_{\gamma})/
(m^{2}_{\beta}-m^{2}_{\gamma})$, $[S^{AI}]_{BJ}=(\sum_{\gamma k}e^{AB\gamma}e^{IJk})\Lambda_{BJ}$, etc.

It is clear that these equations obey the permutation symmetry
$S_{3}(u) \times S_{3}(d)$.  In this connection we recall the 
well-known result \cite{Athanasiu:1986mk} 
about the evolution of $J$, which can be written in the 
form \cite{Chiu:2008ye}  
\begin{equation}\label{III9}
\mathcal{D}\ln [J \cdot \Pi (\Delta m^{2}_{\alpha \beta}) \Pi (\Delta m^{2}_{ij})/\Pi m^{2}_{\alpha}
\Pi m^{2}_{i})]=b(\sum m^{2}_{\alpha}+
\sum m^{2}_{i}).
\end{equation}
Note how $J(\sim \widetilde{\bf{1}} \times \widetilde{\bf{1}})$ combine with
$\Pi(\Delta m^{2}_{\alpha \beta}) \Pi (\Delta m^{2}_{ij}) 
(\sim \widetilde{\bf{1}} \times \widetilde{\bf{1}})$ to form a scalar under
$S_{3}(u) \times S_{3}(d)$, while the other mass combinations in Eq.~(\ref{III9})
are all scalars.

\subsection{A two-loop RGE}

Although most RGE results come from one-loop calculations, there is one two-loop
calculation \cite{Barger:1992ac,Xing:2019tsn} in the literature.  The results, 
written in a way suitable for analyzing their properties under permutation, 
were given in \cite{Chiu:2008ye}.   
Using the same manipulations as in \cite{Chiu:2016qra}, 
we can write the RGE for $W_{\alpha i}$ in the form
\begin{eqnarray}
\mathcal{D}W_{\alpha i}=\mathcal{D}W_{\alpha i}^{(1)} &-&
2c'[\Delta \widetilde{f}'_{\beta}[S^{\alpha i}]_{\beta j}\widetilde{G}_{j}+
\Delta \widetilde{m}^{2}_{\beta}[S^{\alpha i}]_{\beta j}\widetilde{G}'_{j}  \nonumber \\
&+&
\Delta \widetilde{g}'_{j}[S^{\alpha i}]^{T}_{j\beta}\widetilde{H}_{\beta}+
\Delta \widetilde{m}^{2}_{j}[S^{\alpha i}]^{T}_{j\beta}\widetilde{H}'_{\beta}].
\end{eqnarray}
Here, $\mathcal{D}W_{\alpha i}^{(1)}$ denotes the one-loop contribution as 
in Eq.~(\ref{III7}). Without going into details (see Ref.\cite{Chiu:2016qra}), 
the contribution from two-loop is similar in form, except for the introduction
of the primed functions, where $\Delta \widetilde{f}'_{\beta} (\Delta \widetilde{g}'_{k})$
and $\widetilde{G}'_{k} (\widetilde{H}'_{\beta})$ are modified forms of
$\Delta \widetilde{m}^{2}_{\beta} (\Delta \widetilde{m}^{2}_{k})$ and
$G_{k} (H_{k})$, but which transform the same way.
This example suggests that, for any multi-loop calculations,
while the details may differ, one would expect that the result will obey the
permutation symmetry.


\section{ Parametrization of flavor mixing}

Although the SM Lagrangian is given in terms of the mixing matrix $V$,
it contains only four physical variables, and a general problem is the lack of
a criterion to pick an appropriate subset of four parameters amongst those in $V$.
A natural starting point seems to be the use of $W_{\alpha i}$, which are rephasing 
invariant and have clear physical meanings, and try to further pare down the set.
Such a reduction was proposed earlier \cite{Kuo:2005pf}, 
giving rise to a six-parameter set, $(x_{i},y_{j})$, with two consistency conditions.
Permutation symmetry suggests a further reduction, the use of singlets as parameters.
It turns out that, out of the set $(x_{i},y_{j})$, one can construct six singlets,
$\sum x_{i}\pm \sum y_{j}$, $\sum x_{i}x_{j}\pm y_{i}y_{j}$, and
$\Pi x_{i} \pm \Pi y_{j}$, two of which are fixed by the consistency conditions.
(Note that the condition $\sum x_{i}x_{j}-\sum y_{i}y_{j}=const.$ is consistent
with $(\sum x_{i}x_{j}-\sum y_{i}y_{j})$ being a pseudo-P-scalar only if $const.=0$.)
We now propose the following parameter set for flavor mixing,
\begin{equation}
D=\sum x_{i}+\sum y_{j},
\end{equation}
\begin{equation}
Q^{2}=\sum x_{i}x_{j}+\sum y_{i}y_{j},
\end{equation}
\begin{equation}
J^{2}=\Pi x_{i}- \Pi y_{j},
\end{equation}
\begin{equation}
K=\Pi x_{i} +\Pi y_{j}.
\end{equation}
Here, $D$ and $K$ transform as pseudo-P-scalars $(\sim \widetilde{\bf{1}}\times \widetilde{\bf{1}})$,
while $Q^{2}$ and $J^{2}$ are P-scalars $(\sim \bf{1}\times \bf{1})$.  
It can be shown that (see Ref.\cite{Kuo:2005pf} and Appendix B),
$-1 \leq D \leq +1$, $0\leq Q^{2} \leq Q^{2}_{M}$, $0 \leq J^{2} \leq J^{2}_{M}$, and
$-K_{M} \leq K \leq K_{M}$, where the maximal values are given by 
$Q^{2}_{M}=1/6$, $J^{2}_{M}=1/108$, and $K_{M}=2(4/27)^{3}$. The extremal value for $D$
happens when $W=[I]$, which also gives $Q^{2}=0$, $J^{2}=0$, and $K=0$.
$Q^{2}_{M}$ and $J^{2}_{M}$ are attained
with $[W_{M}]$,
\begin{equation}\label{V5}
[W_{M}]=\frac{1}{3}
\left(\begin{array}{ccc}
   1& 1 & 1 \\
   1 & 1 & 1 \\
   1 & 1 & 1 \\
    \end{array}
    \right),
\end{equation}
corresponding to ``maximal mixing". Finally, for $[W]=[W_{K}]$,
\begin{equation}\label{V6}
[W_{K}]=\frac{1}{9}
\left(\begin{array}{ccc}
   1,& 4, & 4 \\
   4, & 1, & 4 \\
   4, & 4, & 1 \\
    \end{array}
    \right),
    \end{equation}
$|K|$ assumes the value $K_{M}$.

\begin{table}
\centering
\begin{tabular}{c||c|c|c|c|c} \hline

 \diaghead(-1,1){aaaaaaa}{}{$[W]$} & $[I]$ &$[W_{M}]$ & $[W_{K}]$ & $W_{AI}=0$& equal row/column \\ \hline \hline
 $D$ & $|D_{M}|=1$ & 0 & 1/9 & ? & 0 \\ \hline
 $Q^{2}$ & 0 & $Q^{2}_{M}=1/6$ & $(2/3)^{5}$ & ?  &  ? \\ \hline
$J^{2}$ &0 & $J^{2}_{M}=2\cdot (1/6)^{3}$ & 0 & 0 & ? \\ \hline
$ K$ & 0 & 0 & $|K_{M}|=2 \cdot (4/27)^{3}$ & 0 & 0 \\ \hline
\end{tabular}
\caption{Values for ($D$, $Q^{2}$, $J^{2}$, $K$) for special $[W]$'s.
$W_{AI}=0$ denotes a matrix $[W]$ with a zero anywhere. Equal row/column indicates that
any two rows or columns are the same. $[W_{M}]$ and $[W_{K}]$ are defined in
Eqs.~(\ref{V5}) and (\ref{V6}).}
\end{table}

The value of $(D,Q^{2},J^{2}.K)$ for various special matrices $[W]$ can be
summarized in Table I. 
Note that $W_{AI}=0$ implies $V_{AI}=0$, so that there is a pair of 
vanishing $(x_{a},y_{b})$, and $J^{2}=K=0$. If there are two zeros in $[W]$,
then consistency  (Eqs.(\ref{con6})-(\ref{con7})) implies that there are at least
four zeros in $[W]$, and the problem reduces to two flavor mixing, with $Q^{2}=J^{2}=K=0$.
The values in the Table remains the same (up to a $\pm$ sign)
when a permutation is made to the matrix $[W]$. 
For instance, instead of the unit matrix $[I]$, any permuted $[I]$,
i.e., the set $\{X\}$ given in Eq.~(\ref{eq17}), would yield the values
$\{D,Q^{2},J^{2},K\}=\{\pm 1,0,0,0\}$.

Note also that, since the variables $x_{i}$ satisfy a cubic equation with coefficients
$(-\sum x_{i},\sum x_{i}x_{j},-\Pi x_{i})$, with a similar equation for $y_{i}$,
knowing $\{D,Q^{2},J^{2},K\}$ would determine both cubic equations.
I.e., one can solve for $(x_{i},y_{j})$ if the parameters $\{D,Q^{2},J^{2},K\}$ are given.
Thus, $\{W_{AI}\}$, $\{x_{i},y_{j}\}$, and $\{D,Q^{2},J^{2},K\}$ are equivalent ways
to describe the same mixing configuration. However, the set $\{D,Q^{2},J^{2},K\}$
is unique in that it does not contain superfluous variables.

We now turn to a discussion of the RGE evolution of the variables $\{D,Q^{2},J^{2},K\}$.
To do that we can use the results of \cite{Chiu:2016qra}, 
where the RGE for $(x_{i},y_{j})$ were obtained ( see Eqs. (\ref{eq26})-(\ref{eq27})):
\begin{eqnarray}
-\mathcal{D}(x_{i})/c'&=&[\Delta \widetilde{m}^{2}_{l}](2[Z_{i}]-[ReZ^{0}])
[\widetilde{G}_{\nu}]^{T} \\ \nonumber
&+& [\Delta \widetilde{m}^{2}_{\nu}](2[Z_{i}]-[ReZ^{0}])^{T}[\widetilde{H}_{l}]^{T},
\end{eqnarray}
\begin{eqnarray}
-\mathcal{D}(y_{i})/c'&=&[\Delta \widetilde{m}^{2}_{l}](2[Z'_{i}]-
[ReZ^{0}])[\widetilde{G}_{\nu}]^{T} \\ \nonumber
&+& [\Delta \widetilde{m}^{2}_{\nu}](2[Z'_{i}]-[ReZ^{0}])^{T}[\widetilde{H}_{l}]^{T}.
\end{eqnarray}
Here, the notation is as in Eqs.(\ref{IV7}-\ref{IV8}), with 
$[\Delta \widetilde{m}^{2}_{l}]=[\Delta \widetilde{m}^{2}_{e},\Delta \widetilde{m}^{2}_{\mu},
\Delta \widetilde{m}^{2}_{\tau}]$, and $[\widetilde{G}_{\nu}]=[\widetilde{G}_{1},\widetilde{G}_{2},\widetilde{G}_{3}]$, etc. The matrices $[Z_{i}]$,$[Z'_{i}]$, and $[ReZ^{0}]$ are given in Table II,
which is adopted from a similar table in Ref. \cite{Chiu:2016qra}, 
written in a notation which is consistent with this paper.


\begin{table}\label{tZ}
 \begin{center}
 \begin{tabular}{ccc}
   \\     \hline \hline  
\vspace{0.15in}   
  $Z_{1}=\left(\begin{array}{ccc}
   \Lambda_{e1} & 0 & 0 \\
    0 & \Lambda_{\mu2} & 0 \\
    0 &0 &\Lambda_{\tau 3} \\
    \end{array}
    \right)$, 
 &  $Z_{2}=\left(\begin{array}{ccc}
   0 & \Lambda_{e2} & 0 \\
    0 & 0 & \Lambda_{\mu3} \\
    \Lambda_{\tau1}& 0 &0 \\
    \end{array}
    \right)$, 
 
    &  $Z_{3}=\left(\begin{array}{ccc}
   0 & 0 & \Lambda_{e3} \\
   \Lambda_{\mu1} & 0 & 0 \\
    0 &\Lambda_{\tau 2} & 0 \\
    \end{array}
    \right)$ 
 
 \\ \vspace{.15in} 
   $Z'_{1}=\left(\begin{array}{ccc}
   \Lambda_{e1} & 0 & 0 \\
   0 & 0 & \Lambda_{\mu 3} \\
    0 &\Lambda_{\tau 2} &0 \\
    \end{array}
    \right)$,    
  &  $Z'_{2}=\left(\begin{array}{ccc}
   0 & \Lambda_{e2} & 0 \\
    \Lambda_{\mu 1} & 0 & 0 \\
   0 & 0 &\Lambda_{\tau 3} \\
    \end{array}
    \right)$,  
   
  &  $Z'_{3}=\left(\begin{array}{ccc}
   0 & 0 & \Lambda_{e3} \\
    0 & \Lambda_{\mu 2} & 0 \\
    \Lambda_{\tau 1} &0 &0 \\
    \end{array}
    \right)$     
   \\  \hline 
   \end{tabular}
  \begin{tabular}{ccc}
\vspace{0.25in}

  $[ReZ^{0}]$ & = & $\left(\begin{array}{ccc}
   x_{2}x_{3}+y_{2}y_{3}, & x_{1}x_{3}+y_{1}y_{3}, & x_{1}x_{2}+y_{1}y_{2}\\
    x_{1}x_{2}+y_{1}y_{3}, & x_{2}x_{3}+y_{1}y_{2}, & x_{1}x_{3}+y_{2}y_{3} \\
   x_{1}x_{3}+y_{1}y_{2},& x_{1}x_{2}+y_{2}y_{3}, & x_{2}x_{3}+y_{1}y_{3} \\
    \end{array}
    \right)$  \\   
   
   \hline \hline
   
    \end{tabular}
 \caption{Explicit expressions for $[Z_{i}]$, $[Z'_{i}]$, and $[ReZ^{0}]$.} 
   \end{center}
 \end{table}  

Using Table II, it is straightforward to find the RGE for $\{D,Q^{2},J^{2},K\}$. The results are
\begin{eqnarray}
\frac{1}{2c'}\mathcal{D}(D)&=&[\Delta \widetilde{m}^{2}_{l}]([\Lambda]-\frac{3}{4}([w^{2}]-[W^{2}]))[\widetilde{G}_{\nu}]^{T} \\ \nonumber
&+&[\Delta \widetilde{m}^{2}_{\nu}]([\Lambda]-\frac{3}{4}([w^{2}]-[W^{2}]))^{T}[\widetilde{H}_{l}]^{T},
\end{eqnarray}
\begin{eqnarray}
\frac{2}{c'}\mathcal{D}(Q^{2})&=&[\Delta \widetilde{m}^{2}_{l}]([(w-D)(w^{2}-W^{2})])[\widetilde{G}_{\nu}]^{T} \\ \nonumber
&+&[\Delta \widetilde{m}^{2}_{\nu}]([(w-D)(w^{2}-W^{2})])^{T}[\widetilde{H}_{l}]^{T},
\end{eqnarray}
\begin{eqnarray}
-\frac{1}{2c}\mathcal{D}(\ln J^{2})&=&[\Delta \widetilde{m}^{2}_{l}][w][\widetilde{G}_{\nu}]^{T} \\ \nonumber
&+& [\Delta \widetilde{m}^{2}_{\nu}][w]^{T}[\widetilde{H}_{l}]^{T},
\end{eqnarray}
\begin{eqnarray}\label{eq66}
-\frac{1}{2c'}\mathcal{D}(K)&=&[\Delta \widetilde{m}^{2}_{l}]([ReZ^{0}(\Lambda-Q^{2}/2)])[\widetilde{G}_{\nu}]^{T} \\ \nonumber
&+&[\Delta \widetilde{m}^{2}_{\nu}]([ReZ^{0}(\Lambda-Q^{2}/2)])^{T}[\widetilde{H}_{l}]^{T}.
\end{eqnarray}
Here, $[w^{2}]$ does not mean matrix multiplication, but indicates products of individual matrix
elements. E.g., $([w^{2}]-[W^{2}])_{AI}=w^{2}_{AI}-W^{2}_{AI}$.
Similarly, $[ReZ^{0}\Lambda]_{AI}=(ReZ^{0})_{AI}\Lambda_{AI}$, etc.

We note that these equations are all of the same structure -- products of three matrices.  
The first and last contain mass matrices and the middle one depends on the mixing parameters.
They are tensor equations under $S_{3}(\ell) \times S_{3}(\nu)$, and are of the form
$[\widetilde{\bf{3}}][\widetilde{\bf{3}}\times \widetilde{\bf{3}}][\widetilde{\bf{3}}]^{T}$ (for $\mathcal{D}(Q^{2})$ and $\mathcal{D}( J^{2})$) and $[\widetilde{\bf{3}}][\bf{3}\times \bf{3}][\widetilde{\bf{3}}]^{T}$
(for $\mathcal{D}(D)$ and $\mathcal{D}(K)$).
Using Eqs.(\ref{eq32}-\ref{eq33}), it is seen that they depend only on $[\Lambda]$ and $[\Lambda']$,
plus the singlets $\{D,Q^{2},J^{2}\}$.  As a consistency check, the variation on any variable
has to vanish when it reaches its maximum or minimum.
For $D$, this happen at $[W]=[X_{i}]$. Indeed, $[\Lambda]=0$ and $[w^{2}]-[W^{2}]=0$ here.
For $Q^{2}$, $[W]=[W_{M}]$ or $[X_{i}]$ implies $w=0$ or $[w^{2}]-[W^{2}]=0$.  
For $J^{2}$, since  $\mathcal{D}(J^{2}) \propto J^{2}$, it vanishes if $J^{2}=0$.
At $J_{M}^{2}$, $[W]=[W_{M}]$, and $[w]=0$. Finally, $K=K_{M}$ at $[W]=[W_{K}]$.
A detailed calculation 
gives $[ReZ^{0}(\Lambda-Q^{2}/2)] \propto [W_{M}]$, with vanishing
contribution to Eq.(\ref{eq66}), as we will verify in Appendix B.

\section{conclusion}
 
 In the SM, the physics of flavor is derived from the mixing matrices and the mass terms
 in the Lagrangian.  Rephasing of the wave functions, which leaves the mass terms intact,
 can be cancelled by corresponding phases applied to $V_{\alpha i}$. This results in
 rephasing invariance, whereby any $V_{\alpha i}$'s which differ by rephasing are 
 physically equivalent. Similarly, if we subject the wave function to a permutation,
 it can be neutralized by a permutation both in $V_{\alpha i}$ and in the masses,
 and we have permutation symmetry. Physically, the validity of this argument is owing to
 the absence of a mechanism in $\mathcal{L}_{SM}$ which can pre-assign the order of
 the families. This freedom to rearrange the family order gives rise to the
 permutation symmetry.
 
 In this paper we emphasize the close analogy between rephasing invariance and permutation
 symmetry. While physical variables are invariants under rephasing (abelian), 
 they are found to be tensors under permutation (nonabelian). A prime example is $J$,
 the Jarlskog invariant, which transforms as a pseudo-P-scalar.
 This assignment is verified in many examples, and offers insights about these relations.
 More generally, the laws of evolution of these physical tensors are shown to obey the permutation symmetry. As tensor equations, they are also much simpler in form than the evolution
 equations written in other variables.
 
 As for the physical applications, we note that the implications of permutation symmetry
 are most noticeable when we consider processes which include internal fermions.
 In this case, the final results do not contain the wave functions, but are functions
 of masses and mixing parameters only. Permutation symmetry then implies that these functions must contain only invariants, such as $f(m_{\alpha}\Lambda_{\alpha i}m_{i})$.
 This is borne out by the examples given in the text.

 The use of symmetry groups in flavor physics has a long history. 
 (For a review, see Ref.\cite{Xing:2019vks}.) Traditionally, they were used to suggest possible
 patterns in the vacuum mass matrix parameters. On the other hand, in this work $S_{3}$ is
 established as the symmetry group of SM Lagrangian. As such $S_{3}$ governs physical transitions and,
 in particular, the evolution equations of the physical mass parameters.  In this sense the vacuum
 parameters are similar to the arbitrary initial values of physical variables in a dynamical system 
 with rotational symmetry. These variables evolve according to the equations of motion, which
 are invariant in form under rotations.
 
 Lastly, permutation properties can be used to identify a unique set of four mixing parameters,
$\{D,Q^{2},J^{2},K\}$. Properties of the set were given.

\acknowledgments                 
We thank Chu-Ching Huang for useful discussions in numerical work.
SHC is supported by the Ministry of Science and Technology of Taiwan, 
Grant No. MOST 107-2119-M-182-002.

\appendix

\section{Relations among permutation tensors}

Physical variables in flavor physics are constructed out of the basic parameters
$(m_{\alpha},m_{i},V_{\alpha i})$. From the masses, simple functions such as
$\widetilde{m}^{2}_{\alpha}=\frac{1}{2}e_{\alpha \beta \gamma}(m^{2}_{\beta}-m^{2}_{\gamma})$
are obtained and used. These results are straightforward and no further elaborations
are necessary. The situation is more complex for $V_{\alpha i}$. Here, physical variables
are built up from basic RIC blocks given by $\Gamma_{IJK}^{AB\Gamma}$ (Eq.~(\ref{II4})),
which consists of a triplet of $V_{\alpha i}$'s and is characterized by having the index of
each family once, and only once ($V^{*}_{\alpha i}$ is to be considered as the product
of two $V_{\beta j}$'s, by Eq.~(\ref{II2}).)   
As was mentioned before, any permutation of the indices of $\Gamma$ yields another RIC.
Starting from any such combination, repeated permutations generate a set of physical variables
which form a tensor. With the only available tensors in $S_{3}$ being singlets or triplets,
we can anticipate that, out of the myriad possible ways to put together the $V_{\alpha i}$'s
into tensors, there are only a few independent ones. In this Appendix, we will summarize
the identities connecting the various tensors coming from seemingly different constructions.

Several of these identities were already given in Eqs.~(\ref{eq11}-\ref{eq16})
and Eqs.~(\ref{eq26}-\ref{eq33}).  We now write down more which are useful in our
calculations.  We begin by giving the tensors $\Lambda_{AI}$ and $\Lambda'_{AI}$
in another useful form
\begin{equation}
\Lambda_{AI}=[Re Z^{0}]_{AI}-\frac{Q^{2}}{2}+\frac{1}{4}(w^{2}_{AI}-W^{2}_{AI}),
\end{equation}
\begin{equation}
\Lambda'_{AI}=[Re Z^{0}]_{AI}-\frac{Q^{2}}{2}-\frac{1}{4}(w^{2}_{AI}-W^{2}_{AI}).
\end{equation}
These relations can be verified by using Eqs.~(\ref{eq26}-\ref{eq27}) by writing the variables
in terms of $(x_{i},y_{j})$. Similarly, we can establish the following identities:
\begin{eqnarray}
4[Re Z^{0}]_{AI}&=&E_{A \beta \gamma}(W_{\beta I}W_{\gamma I}+w_{\beta I}w_{\gamma I}) \\ \nonumber
&=& E_{Ijk}(W_{Aj}W_{Ak}+w_{Aj}w_{Ak})
\end{eqnarray}
\begin{equation}
W_{AI}\Lambda_{AI}=J^{2}+\frac{1}{4}(w^{2}_{AI}-W^{2}_{AI}),
\end{equation}
\begin{equation}
W_{AI}\Lambda'_{AI}=J^{2}-\frac{1}{4}(w^{2}_{AI}-W^{2}_{AI})(2W_{AI}-1).
\end{equation}
All of these composite terms 
$(\Lambda_{AI},\Lambda'_{AI},[Re Z^{0}]_{AI},(w^{2}_{AI}-W^{2}_{AI}), ....)$ transform as $\bf{3}\times \bf{3}$ and/or $\bf{1}\times \bf{1}$, but there are only two independent
$\bf{3}\times \bf{3}$ (e.g., $\Lambda_{AI},\Lambda'_{AI}$) plus two $\bf{1} \times \bf{1}$
($J^{2},Q^{2}$) tensors.

Similarly, we can construct composite tensors which transform as 
$\widetilde{\bf{3}} \times \widetilde{\bf{3}}$ and $\widetilde{\bf{1}} \times \widetilde{\bf{1}}$.
These include the tensor 
$[\Xi^{0}]_{AI} (\sim \widetilde{\bf{3}} \times \widetilde{\bf{3}}+\widetilde{\bf{1}} \times \widetilde{\bf{1}})$, defined in Eq.~(\ref{eq28}). It can be given in another form
\begin{eqnarray}
2[Re \Xi^{0}]_{AI}&=&-E_{A \beta \gamma}W_{\beta I}w_{\gamma I} \\ \nonumber
&=&-E_{Ijk}W_{Aj}w_{Ak}.  \\ \nonumber
\end{eqnarray}
In addition,
\begin{equation}
[Re \Xi^{0}]_{AI}\Lambda_{AI}=-J^{2}w_{AI}+\frac{1}{2}(K+DJ^{2}),
\end{equation}
which was used in writing down the RGE of $J^{2}$.
Finally, we give two more identities which transform as 
$\widetilde{\bf{3}} \times \widetilde{\bf{3}}+\widetilde{\bf{1}} \times \widetilde{\bf{1}}$:
\begin{equation}
w_{AI}\Lambda_{AI}=K+\frac{1}{4}(2w_{AI}-D)(w^{2}_{AI}-W^{2}_{AI}),
\end{equation}
\begin{equation}
w_{AI}\Lambda'_{AI}=K-\frac{1}{4}D(w^{2}_{AI}-W^{2}_{AI}).
\end{equation}

\section{Properties of $D$, $Q^{2}$,$J^{2}$,$K$}

In Sec. IV, we proposed to use the set $\{D,Q^{2},J^{2},K\}$ as parameters for flavor mixing. 
The characteristic of this set is that its members are rephasing invariant, in addition to being
singlets under permutation. These properties are in conformity with $\mathcal{L}_{SM}$,
which is rephasing invariant and exhibits permutation symmetry. 
In this Appendix we discuss some detailed properties of this set.

We start by recalling the properties of the variables $(x_{i},y_{j})$,
which are rephasing invariant but behave as mixed tensors under permutation.
Under simple exchanges ($(\alpha \beta)$ or $(ij)$), $x_{a} \leftrightarrow -y_{b}$,
while for cyclic permutations, $x_{a} \leftrightarrow x_{c}$ and $y_{b} \leftrightarrow y_{d}$,
for some indices ($a,b,c,d$). But the combinations ($\sum x_{i},\sum y_{j}$, etc.) have simpler
transformation properties, resulting in $\{D,K\} \leftrightarrow \{\pm D,\pm K\}$ and
$\{Q^{2},J^{2}\} \leftrightarrow \{Q^{2},J^{2}\}$, for any permutations.

In Ref.\cite{Kuo:2005pf}, the ranges of $D$ and $J^{2}$ were established.
It was found that $-1 \leq D \leq +1$, where the boundary values are reached when
$[W]=[X]$, which are given in Eq.(\ref{eq17}), and consist of the identity matrix $[I]$
and its permutations. Similarly, $0 \leq J^{2} \leq J^{2}_{M}$, where the maximum,
$J^{2}_{M}=1/108$ is attained with $[W]=[W_{M}]$.

  \begin{figure}[ttt]
\caption{With random choices of the $(x_{i},y_{j})$ values, the occurrence frequencies of values of $Q^{2}$ are plotted using $1.44 \times 10^{5}$ data points generated.} 
\centerline{\epsfig{file=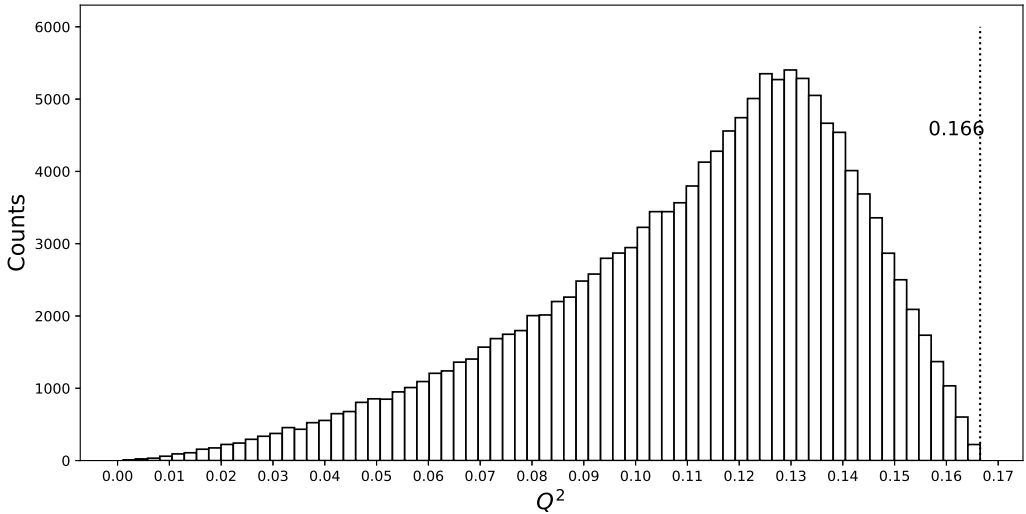,width= 16cm}}
\end{figure} 

  \begin{figure}[ttt]
\caption{With random choices of the $(x_{i},y_{j})$ values, the occurrence frequencies of values of $K$ are plotted using $1.44 \times 10^{5}$ data points generated.} 
\centerline{\epsfig{file=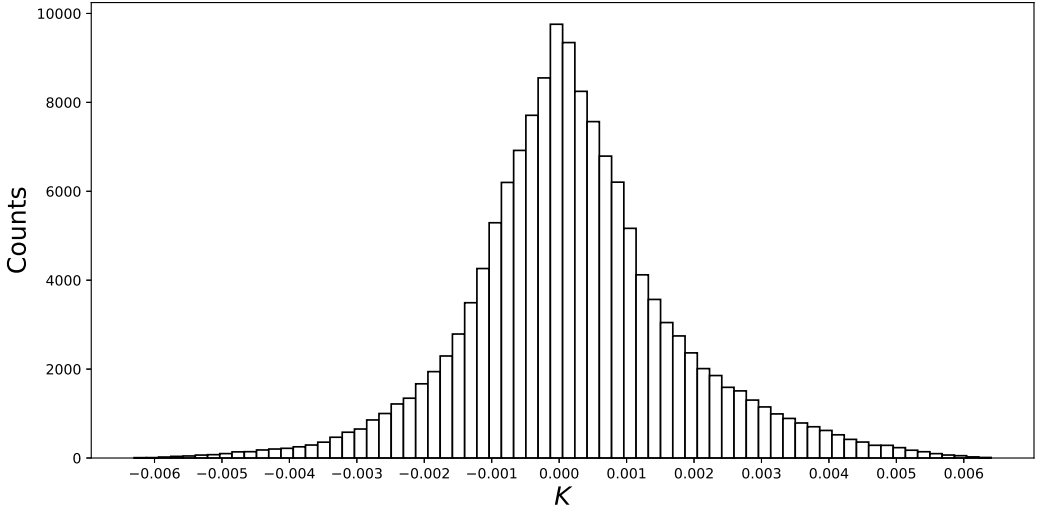,width= 16cm}}
\end{figure}

The ranges of $Q^{2}$ and $K$ are less clear. It was known \cite{Kuo:2005pf} that
$Q^{2} \geq 0$, but we need to find $Q^{2}_{M}$. For the pseudo-P-scalar $K$,
we have to determine $K_{M}$, with $-K_{M} <K<K_{M}$. For lack of an
analytic solution, we resort to a numerical search.
In Figs.1 and 2, we plot the number of occurrences of values of $Q^{2}$ and $K$,
for random choices of the $(x_{i},y_{j})$ values. The result is that $Q^{2}_{M} \approx 0.166$
at $(x_{i};y_{j}) \approx(0.169,0.163,0.168;-0.162,-0.174,-0.164)$ and
$K_{M} \approx 0.006$ at $(x_{i};y_{j}) \approx(-0.037,0.304,0.289;-0.154,-0.148,-0.143)$.  
These values suggest the matrices $[W_{M}]$ and $[W_{K}]$, Eqs. (\ref{V5}-\ref{V6}),
to be the exact locations for the maxima of $Q^{2}$ an $|K|$.
The $(x,y)$ values are $x_{i}=-y_{j}=1/6$ for $[W_{M}]$ and
 $(x_{i};y_{j}) = (-1/27,8/27,8/27;-4/27,-4/27,-4/27)$ for $[W_{K}]$.
 Using these values, this conjecture can be verified as follows,
 Starting from any given $(x_{i},y_{j})$, we consider the variation 
 $(x_{i}+\delta x_{i},y_{j}+\delta y_{j})$. To satisfy the consistency conditions
 (Eqs.(\ref{con6})-(\ref{con7})), we have
 \begin{equation}
 \sum \delta x_{i}-\sum \delta y_{j}=0,
 \end{equation}
\begin{equation}
\sum_{i\neq j}x_{i} \delta x_{j}=\sum_{\j \neq j} y_{i} \delta y_{j}.
\end{equation}

At $[W_{M}]$, $x_{i}=-y_{j}=1/6$, these conditions imply $\sum \delta x_{i}=\sum \delta y_{i}=0$.
It follows that 
\begin{eqnarray}
\delta (Q^{2}/2) &=& \sum_{i<j}[(x_{i}+\delta x_{j})(x_{j}+\delta x_{j})-x_{i}x_{j}] \\ \nonumber
&=& \frac{1}{3}\sum(\delta x_{i})+\sum_{i<j}(\delta x_{i})(\delta x_{j}) \\ \nonumber
&=&-\frac{1}{2} \sum (\delta x_{i})^{2}.
\end{eqnarray}
This verifies that $Q^{2} \leq Q^{2}_{M}$, with $Q^{2}_{M}=1/6$.

At $[W_{K}]$, Eq.(\ref{V6}), $(x_{i};y_{j}) = (-1/27,8/27,8/27;-4/27,-4/27,-4/27)$.
The consistency conditions are now $\sum \delta x_{i}=\sum \delta y_{j}$,
$5\sum \delta x_{i}+3 \delta x_{1}=0$. But at $[W_{K}]$, $J^{2}=0$, which
is at the boundary of $J^{2}$ so that also $\delta J^{2}=0$.
These gives $3\delta x_{1}-\sum \delta x_{i}=0$. Taken together, we find that at $[W_{K}]$,
the constraints on variations of ($x_{i},y_{j}$) are 
$\sum \delta x_{i}=\sum \delta y_{i}=\delta x_{1}=0$ (and $\delta x_{2}+\delta x_{3}=0$).
Now, at $[W_{K}]$, we have
\begin{eqnarray}
\delta K &=& \Pi (x_{i}+\delta x_{i})+\Pi (y_{i}+\delta y_{i})-\Pi x_{i} -\Pi y_{i} \\ \nonumber
&=& (\delta x_{2})^{2}+\frac{1}{2} \sum (\delta y_{i})^{2}+ \mathcal{O}(\delta^{3}).
\end{eqnarray}
I.e., $K$ is at a minimum for $K=-K_{M}$ at $[W]=[W_{K}]$.

The results above are summarized in Table I.  Finally, to verify that $\mathcal{D}(K)=0$
at $|K|=K_{M}$, we note that, with $[W]=[W_{K}]$,
\begin{equation}
[\Lambda_{K}]=\frac{4}{81}
\left(\begin{array}{rrr}
   1,& -2, & -2 \\
   -2, & 1, & -2 \\
   1, & -2, & 1 \\
    \end{array}
    \right),
    \end{equation}
\begin{equation}
[ReZ^{0}_{K}]=\frac{8}{(27)^{2}}
\left(\begin{array}{rrr}
   10,& 1, & 1 \\
   1, & 10, & 1 \\
   1, & 1, & 10 \\
    \end{array}
    \right).
    \end{equation}
With $Q^{2}_{K}=(2/3)^{5}$, this shows that 
$[ReZ^{0}(\Lambda-\frac{Q^{2}}{2})]_{K}=-\frac{5 \cdot (2)^{6}}{(3)^{10}}\cdot W_{M}$,
whose contribution to Eq.(\ref{eq66}) indeed vanishes.

\end{document}